\documentstyle[11pt,newpasp,twoside,epsf]{article}
\def\spose#1{\hbox to 0pt{#1\hss}}
\def\lta{\mathrel{\spose{\lower 3pt\hbox{$\mathchar"218$}}
     \raise 2.0pt\hbox{$\mathchar"13C$}}}
\def\gta{\mathrel{\spose{\lower 3pt\hbox{$\mathchar"218$}}
     \raise 2.0pt\hbox{$\mathchar"13E$}}}
\def\edcomment#1{\iffalse\marginpar{\raggedright\sl#1\/}\else\relax\fi}
\reversemarginpar

\begin{document}
\title{Young Stellar Groups, Runaway Stars, and Pulsars} 
\author{Tim de Zeeuw, Ronnie Hoogerwerf \& Jos de Bruijne} 
\affil{Leiden Observatory, Postbus 9513, 2300 RA Leiden, The Netherlands}

\begin{abstract}
Milli-arcsecond astrometry provided by Hipparcos and by radio
observations makes it possible to retrace the orbits of nearby runaway
stars and pulsars with sufficient accuracy to identify their parent
stellar cluster or association. For two cases it is even possible to
deduce the specific formation scenario. The runaway star $\zeta$ Oph
and PSR J1932+1059 are the result of a supernova explosion which took
place 1 Myr ago in a massive binary in the Upper Scorpius
association. The pulsar received a kick velocity of about 350 km
s$^{-1}$ in this event. The runaway stars $\mu$ Col and AE Aur and the
isolated eccentric massive binary $\iota$ Ori result from a
binary-binary encounter, most likely inside the Trapezium cluster, 2.5
Myr ago. Future astrometric missions such as DIVA, FAME and in
particular GAIA will allow extension of these studies to a significant
fraction of the Galactic disk, and will provide new constraints on the
formation and evolution of massive stars.
\end{abstract}

\section{Introduction}

Most O and B stars in the Galactic disk are found in young star
clusters or OB associations, where they often reside in
binaries. About 30\% of the O stars and 5--10\% of the B stars are
found in the field. These objects almost invariably have large space
velocities (30--200 km s$^{-1}$). Most of these `runaway' stars appear
to be single. Many have high He abundance, and rotate fast.

Two scenarios for the formation of runaway stars are considered
viable: (i) a supernova explosion in a massive binary (Zwicky 1957;
Blaauw 1961), in which the primary becomes a compact object, and the
secondary moves away with a velocity comparable to its pre-explosion
orbital velocity, and (ii) dynamical ejection from a dense stellar
cluster (Poveda et al.\ 1967; Gies \& Bolton 1986). Both scenarios can
explain the properties of the ensemble of runaway stars observed to
date. As a result, the relative importance of the two scenarios has
been debated extensively in the past decades.
 
In order to find specific rather than statistical evidence for either
scenario, we need to establish a common origin for either a runaway
star and a pulsar or a runaway star and a dense stellar cluster. The
availability of Hipparcos milli-arcsec (mas) global astrometry for the
nearby stars, and of pulsar astrometry provided by VLBI and timing
measurements, therefore stimulated us to carry out a systematic study
of the past orbits of the nearby runaways and pulsars. The full study
will be reported elsewhere (Hoogerwerf et al.\ 2000); here we
restrict ourselves to a brief summary of some of the main results.

\newpage
\begin{figure}[t]
\plotfiddle{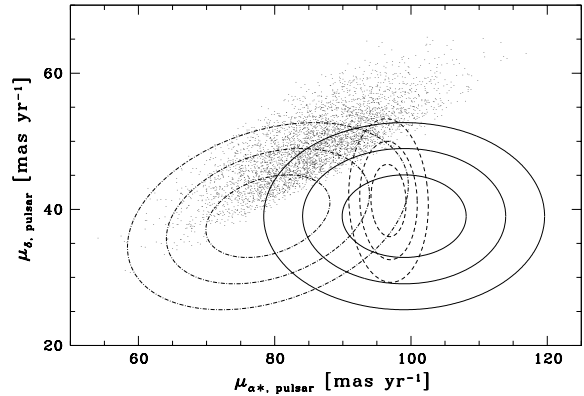}{4.5truecm}{0.0}{40.0}{40.0}{-132.0}{-132.0}
\caption{Measurements of the proper motion of PSR J1932+1059. Dot-dash
curves: Lyne, Anderson \& Salter (1982), solid curves: Taylor,
Manchester \& Lyne (1993), and dashed curves: Campbell (1995).
Contours indicate the 1, 2, and 3$\sigma$ confidence levels. The small
dots are the values consistent with a close encounter between $\zeta$
Oph and PSR J1932+1059 about 1 Myr ago. }
\end{figure}

\begin{figure}[t]
\plotfiddle{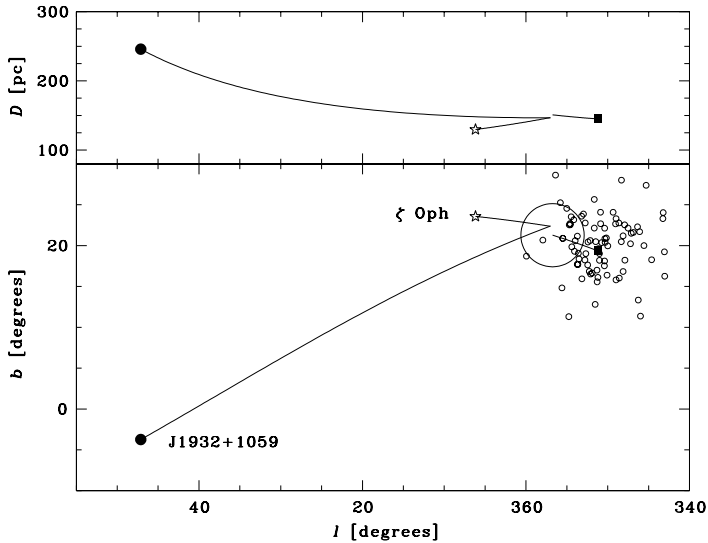}{6.5truecm}{0.0}{48.0}{48.0}{-152.0}{-86.0}
\caption{The orbits of $\zeta$ Oph, PSR J1932+1059, and Upper
Scorpius. The present positions are denoted by a star, a filled
circle, and a filled square (center of the subgroup). Top: distance
$D$ vs Galactic longitude $\ell$. Bottom: orbits projected on the
plane of the sky in Galactic coordinates. The present-day positions of
the O, B, and A-type members of Upper Scorpius are indicated by the
small circles (de Zeeuw et al.\ 1999). The large circle indicates the
position of the subgroup when the supernova explosion created the
pulsar and the runaway star. }
\end{figure}

\section{A binary supernova in Upper Scorpius}

Blaauw (1952) proposed that the O9.5V runaway star $\zeta$ Oph
originated in the nearby association Sco OB2. It is a single star,
appears to be a blue straggler, and has He abundance $Y\sim 0.40$ and
$v_{\rm rot}\sin i=348$ km s$^{-1}$. The space motion of $\zeta$ Oph
is $\sim$40 km s$^{-1}$ relative to Sco OB2, and the star traversed
the 5 Myr old subgroup Upper Scorpius $\sim$1 Myr ago. The kinematics
of the expanding HI shell which surrounds Upper Scorpius is consistent
with a supernova explosion around the same time, and the present-day
mass function of the subgroup indicates that about 40 $M_{\sun}$ is
missing (de Geus 1992). It is therefore natural to speculate that the
supernova occurred in a massive binary, and that $\zeta$ Oph is the
former secondary. Mass transfer during the binary evolution prior to
the explosion would then be the cause of the increased He abundance,
the large rotation velocity, and the blue straggler nature. Can we
find the compact object?

The catalog of pulsars with distances and proper motions maintained by
Taylor, Manchester \& Lyne (1993) contains seven objects within 1 kpc
with measurement accuracy better than 10\%. The radial velocities
$v_{\rm rad}$ of these pulsars are unknown, so we retraced their paths
in the gravitational field of the Galaxy by numerical orbit
integration for a range of $\pm$ 500 km s$^{-1}$ in $v_{\rm
rad}$. Taking into account the characteristic age $P/2{\dot P}$ of the
pulsars leaves only one likely candidate: PSR J1932+1059.  If its
$v_{\rm rad}=200\pm50$ km s$^{-1}$, then it passed through Upper
Scorpius about 1 Myr ago, on its way towards the Galactic plane (which
it recently crossed). If the pulsar did not originate in Upper
Scorpius, then it must have been launched from the Galactic plane some
$\sim$ 50 Myr ago. This is unlikely given a characteristic age of
about 3 Myr.

Having identified the most likely pulsar, we computed the absolute
minimum separation $D_{\rm min}$ between the orbits of $\zeta$ Oph and
PSR J1932+1059, while sampling the error distributions in position and
velocity. The main uncertainties are the errors in the parallax of
$\zeta$ Oph ($\pi=7.1\pm0.7$ mas) and the pulsar ($\pi=5\pm1.5$ mas,
Campbell 1995), and the remaining range in $v_{\rm rad}$.  We found
that the distribution of $D_{\rm min}$ is consistent with $\zeta$ Oph
and the pulsar being {\it in the same location} $1.0\pm0.1$ Myr ago in
Upper Scorpius (Figures 1 \& 2). This is compelling dynamical evidence
for the binary supernova scenario.

The value of $P/2\dot P$ is an uncertain age indicator, and the
$\sim3$ Myr for PSR J1932+1059 is consistent with the dynamical age of
1 Myr derived here. While the current period is 0.22 seconds, the
implied period at birth is 0.18 seconds.

Pulsars are expected to receive a kick velocity ${\vec v}_{\rm kick}$
at birth. Observations of the ensemble of pulsars suggest that ${\vec
v}_{\rm kick}$ is a few hundred km s$^{-1}$ (Hartman 1997; Hansen \&
Phinney 1997). Assuming that $\zeta$ Oph and PSR J1932+1059 originated
in the same binary allows a determination of ${\vec v}_{\rm kick}$.
Based on the angle between the orbits of the two objects, and the
magnitude of the space velocities, we find $|{\vec v}_{\rm kick}| =
350\pm50$ km s$^{-1}$. New VLBI observations of the proper motion and
parallax of PSR J1932+1059 which are being obtained by Campbell
(priv.\ comm.) will make it possible to improve this estimate further.

\begin{figure}
\plotfiddle{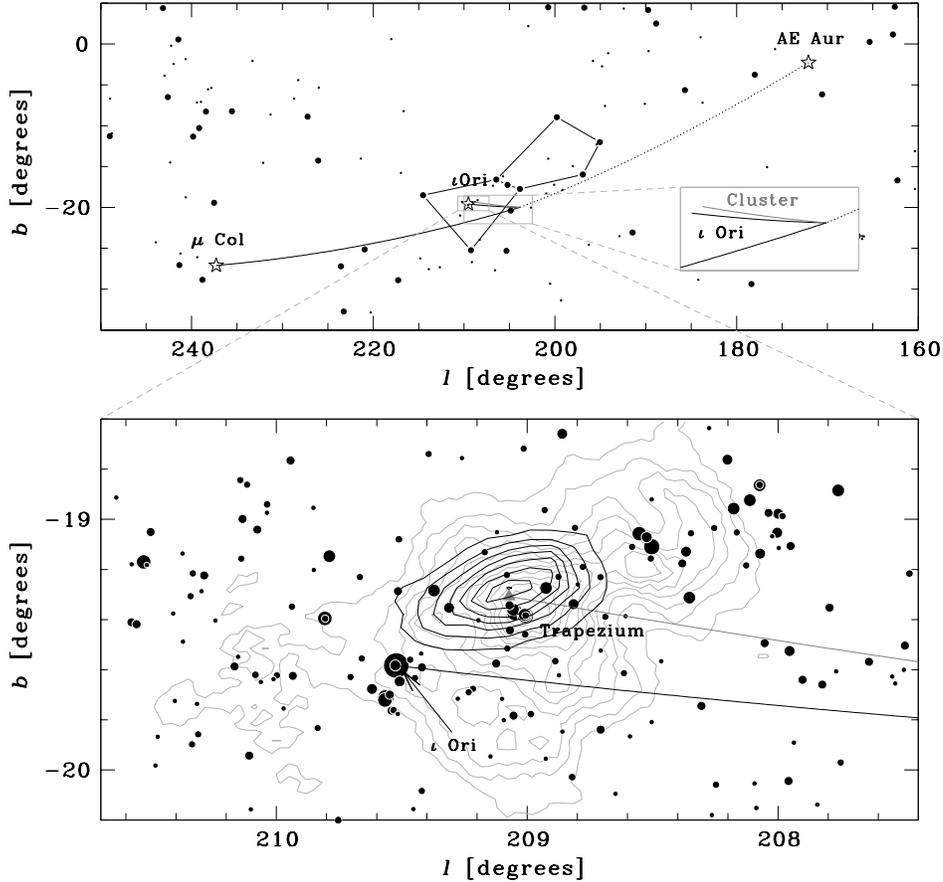}{11.0truecm}{0.0}{64.0}{64.0}{-196.0}{-86.0}
\caption{The orbits of $\mu$ Col, AE Aur and the binary $\iota$ Ori
projected on the sky. All three had a simultaneous encounter 2.5 Myr
ago. The grey line is the orbit of the parent cluster since the
encounter. The large dots denote the stars in the Hipparcos Catalog
brighter than $V=3.5$ mag. The small dots denote the O and B type
stars with 3.5 mag $\leq V \leq 5$ mag (cf.\ figure 1 of Morgan \&
Blaauw 1954). The Orion constellation is indicated for reference.
Bottom: The predicted position of the parent cluster (solid contours)
together with all stars in the Tycho Catalog in the field to $V=$ 12.4
mag. The size of the symbols scales with magnitude. The brightest star
is $\iota$ Ori. The dark and grey lines are the past orbits of $\iota$
Ori and the Trapezium, respectively (see top panel). The triangle
indicates the predicted present-day position of the parent cluster.
The grey contours display the IRAS 100$\mu$m flux map, and mainly
outline the Orion Nebula.  }
\end{figure}

%\newpage
\section{A stellar encounter in Orion}

Blaauw \& Morgan (1954) drew attention to the isolated stars AE Aur
(O9.5 V) and $\mu$ Col (O9.5V/B0V), which run away in opposite
directions from the association Ori OB1 with space velocities of about
100 km s$^{-1}$ each (Figure 3). Blaauw \& Morgan speculated that both
stars originated in the same event in Ori OB1, some 2.5 Myr ago.  Gies
\& Bolton (1986) suggested that this event was an encounter between
two hard binaries, which also produced the eccentric massive binary
$\iota$ Ori (O9III+B1III).

We used the globally accurate Hipparcos proper motions and parallaxes,
and published radial velocities to retrace the orbits of $\iota$ Ori
and the two runaways. We carried out $2.5\times 10^6$ experiments to
sample the error distribution in the measurements (the largest
uncertainties are in the parallax, as the objects are at distances of
$\gta 450$ pc). We found that the distribution of the radius $D_{\rm
min}$ of the minimum volume enclosing the three objects is consistent
with AE Aur, $\mu$ Col and $\iota$ Ori being {\it simultaneously in
one location} $2.5\pm0.2$ Myr ago. This is strong evidence for the
dynamical ejection scenario.

It is of interest to ask, where did the binary-binary encounter occur?
We assumed that the center-of-mass motion of the four stars involved
in the encounter was identical to the mean motion of the parent
cluster. By using conservation of linear momentum, we then computed
the current location and mean motion of the cluster. This is somewhat
sensitive to the masses of AE Aur and $\mu$ Col, but all data are
consistent with the binary-binary encounter taking place in the
nascent Trapezium cluster (Figure 3). This identification is supported
by the age of $\sim$2 Myr (Palla \& Stahler 1999), the high stellar
density, the short dynamical time as evidenced by mass segregration
(Hillenbrand \& Hartmann 1998), and the high binary fraction (Weigelt
et al.\ 1999) of the Trapezium cluster.

\section{Conclusions and prospects} 

The results from the previous two sections demonstrate that runaway
stars are created both by binary supernova explosions and by dynamical
ejection, and that it is now possible to identify individual events. 

Enlarging the sample would provide a probe of the recent history of
massive star formation in the Galactic disk, and would constrain the
fraction of binaries in young clusters (Portegies Zwart 1999). We
therefore took all 3622 O--B5 stars in the Hipparcos Catalog with (i)
measured radial velocity $v_{\rm rad}$ (1118), (ii) parallax
$\pi>2\sigma_\pi$ (which translates to a distance less than about 700
pc), (iii) accurate proper motion $\sigma_\mu/\mu <0.1$, and (iv)
space velocity ${\vec v} > 30$ km s$^{-1}$.  This produced 56 runaway
candidates.  Similarly, we took all the pulsars in the Taylor et al.\
(1993) catalog with (i) measured distance and proper motions (94), a
distance estimate of $D<1$ kpc, and (iii) accurate proper motion
$\sigma_\mu/\mu <0.1$. This leaves only seven pulsars. We then
retraced the orbits of all these objects, to identify the sites of
origin. This work is in progress.

Both samples are severely incomplete. The Hipparcos Catalog is
complete only for $V < 7.3-9$ (depending on latitude), and less than a
third of the O and B stars in it have a measured radial velocity.
Because of beamed radio emission, we cannot observe all pulsars, and
not all of those that do radiate in our direction have been found; of
these, only a few have an accurately measured proper motion. Even so,
to date we have found the parent stellar group for two more pulsars
(in addition to PSR J1932+1059) and for $\sim$20 runaway stars (10 of
which are new). The two pulsars both originate in the $\sim 50$ Myr
old association Per OB3. This is not surprising, as the peak of pulsar
production occurs for the lower-mass B stars, i.e., in old groups.
Twenty of the 56 runaway stars can be linked to the nearby
associations.  Other runaways in the sample presumably originate in
open clusters, including a newly-discovered pair of stars running away
in opposite directions from the region of the $\lambda$ Ori
cluster. Further progress requires accurate space motions for the
young open clusters in the Solar neighbourhood.  Some of the remaining
runaways (and pulsars) must have originated beyond 700 pc, where our
knowledge of the parent groups is poor.%\looseness=-2

Much can be learned by measuring radial velocities for the nearby O
and B field stars, and by increasing the sample of pulsars with
measured astrometry. At the current accuracy of about 1 mas in
parallax and 1 mas/yr in proper motion, a full census is possible of
the nearby 500 pc. Extension to a larger volume requires $\mu$as
astrometry, which will be provided by future space observatories such
as DIVA, FAME, and GAIA.

\acknowledgments It is a pleasure to thank A.\ Blaauw, R.\ Campbell
and M.A.C.\ Perryman for stimulating conversations.

\subsubsection{Pavel Kroupa:} 
1) Dynamical ejection from the Trapezium cluster $\approx$2.5 Myr
gives an {\em important} alternative age estimate that is larger than
the typically quoted age.  2) My N-body calculations show that up to
50\% of all stars with $M>8M_{\sun}$ are found at $R>2R_{\rm tidal}$
after a few Myr.  3) Ejected low-mass binaries will have highly
eccentric orbits. This allows for a test of tidal circularisation
theory of fully convective stars.


\begin{references}
\baselineskip 11pt
\reference Blaauw, A. 1952, BAN, 11, 414 
\reference Blaauw, A., \& Morgan, W. W. 1954, \apj, 119, 625 
\reference Campbell, R. M. 1995, PhD Thesis, Harvard University
\reference de Geus, E. J. 1992, \aap, 262, 258
\reference Gies, D. R., \& Bolton, C. T. 1986, \apjs, 61, 419 
\reference Hansen, B. M. S., \& Phinney, E. S. 1997, \mnras, 291, 569 
\reference Hartman, J. W. 1997. \aap, 322, 127 
\reference Hillenbrand, L. A., \& Hartmann, L. W. 1998, \apj, 492, 540
\reference Hoogerwerf, R., de Bruijne, J. H. J., \& de Zeeuw, P. T. 2000, 
           \aap, submitted
\reference Lyne, A. G., Anderson, B., \& Salter, M. J. 1982, \mnras, 201, 503  
\reference Palla, F., \& Stahler, S. W. 1999, \apj, 525, 772
\reference Portegies Zwart, S. 2000, preprint [astro-ph/0005245]
\reference Poveda, A., Ruiz, J., \& Allen, C. 1967, Bol Obs 
           Tonantzintla, 28, 86
\reference Taylor, J. H., Manchester, R.N., \& Lyne, A.G. 1993, \apjs, 88, 529
\reference Weigelt, G., Balega, Y., Preibisch, T., Schertl, D., Sch\"oller, 
           M., \& Zinnecker, H. 1999, \aap, 347, L15           
\reference de Zeeuw, P. T., Hoogerwerf, R., de Bruijne, J. H. J., 
           Brown, A. G. A., \& Blaauw, A. 1999, \aj, 117, 354
\reference Zwicky, F. 1957, Morphological Astronomy, Springer--Verlag, Berlin 
\end{references}
\end{document}